\documentclass[aps,prl,twocolumn,amsmath,amssymb,notitlepage,noeprint,nofootinbib,superscriptaddress]{revtex4-1}
\pdfoutput=1
\usepackage[latin9]{inputenc}
\usepackage{braket}
\usepackage{xcolor}
\usepackage{verbatim}
\usepackage{graphicx}
\usepackage{esint}
\usepackage{epstopdf}
\usepackage{float}
\definecolor{darkblue}{rgb}{0,0,0.6}
\usepackage[colorlinks,linkcolor=darkblue,citecolor=darkblue,urlcolor=darkblue]{hyperref}

\let \D=\Delta

\def\SI{SI}

\begin{document}

\title{Depletion of two-level systems in ultrastable computer-generated glasses}

\author{Dmytro Khomenko*}

\affiliation{Laboratoire de Physique de l'Ecole Normale Sup\'erieure, ENS, Universit\'e PSL, CNRS, Sorbonne Universit\'e, Universit\'e de Paris, 75005 Paris, France
}

\affiliation{Department of Chemistry, Columbia University, New York, NY 10027, USA}

\author{Camille Scalliet*} 
\email{cs2057@cam.ac.uk}

\affiliation{
DAMTP, Centre for Mathematical Sciences, University of Cambridge, Wilberforce Road,
Cambridge CB3 0WA, United Kingdom}

\author{Ludovic Berthier}  

\affiliation{Department of Chemistry, University of Cambridge, Lensfield Road, Cambridge CB2 1EW, United Kingdom}

\affiliation{Laboratoire Charles Coulomb (L2C), Universit\'e de Montpellier, CNRS, 34095 Montpellier, France}

\author{David R. Reichman}

\affiliation{Department of Chemistry, Columbia University, New York, NY 10027, USA}

\author{Francesco Zamponi}  

\affiliation{Laboratoire de Physique de l'Ecole Normale Sup\'erieure, ENS, Universit\'e PSL, CNRS, Sorbonne Universit\'e, Universit\'e de Paris, 75005 Paris, France
}

\let\thefootnote\relax\footnotetext{*Equal contribution to the work}

\date{\today}

\begin{abstract}
Amorphous solids exhibit quasi-universal low-temperature anomalies whose origin has been ascribed to localized tunneling defects. Using an advanced Monte Carlo procedure, we create {\it in silico} glasses spanning from hyperquenched to ultrastable glasses.
Using a multidimensional path-finding protocol, we locate tunneling defects with energy splittings smaller than $k_{B}T_Q$, with $T_Q$ the temperature below which quantum effects are relevant ($T_Q \approx 1 \,$K in most experiments). We find that as the stability of a glass increases, its energy landscape as well as the manner in which it is probed tend to deplete the density of tunneling defects, as observed in recent experiments. We explore the real-space nature of tunneling defects, and find that they are mostly localized to a few atoms, but are occasionally dramatically delocalized.
\end{abstract}

\maketitle

The theory of low-temperature properties of perfect crystals stands as one of the most profound early tests of the power of quantum statistical mechanics.  In particular, Debye's calculation of the observed $T^{3}$ behavior of the low-temperature specific heat highlighted the importance of long wavelength phonons as low energy excitations in ordered solids~\cite{AM76}.  Given that the wavelength of populated phonon modes around $T \sim 1 \,$K is significantly longer than the interparticle distance in a solid, it came as a major surprise in 1971 when Zeller and Pohl~\cite{zellerpohl} measured large deviations from the expected Debye behavior of the specific heat and the thermal conductivity of vitreous silica, selenium and germanium-based glasses.  An explanation for this puzzling observation was almost immediately put forward by Anderson, Halperin and Varma~\cite{anderson72}, and Phillips~\cite{phillips1972}. They posited that the disorder intrinsic to amorphous solids causes their energy landscape to have many minima. Rare, nearly degenerate, adjacent local minima support tunneling defects or two-level systems (TLS) with energy splittings of the order of $1 \,$K, which provide a large excess contribution to the specific heat and a new mode of scattering that determines the thermal conductivity. In the subsequent decades, the behavior described by Zeller and Pohl was observed in numerous other amorphous materials, 
and the TLS theory has withstood essentially all experimental tests~\cite{loponen1982,phillips87,BM88,boiron1999,burin2013}.  Despite this great progress, the microscopic real-space structure of the tunneling defects remains debated, 
as do the factors that determine their density and distribution in amorphous solids~\cite{galperin1989,coppersmith1991,leggett1991,burin1996,reichman1996,lubchenko2001,gurarie2003,lubchenko2003,parshin2007,vural2011,leggett2013,zhou2015,carruzzo2019}.

A powerful platform for addressing these issues is the use of computer simulation to prepare amorphous materials {\it in silico} and to interrogate the simulated energy landscape for TLS~\cite{He08}. 
This program was initiated by Stillinger and Weber~\cite{SW82,SW85}, then carried out more extensively by Heuer and Silbey~\cite{heuer1993microscopic,heuer1994tunneling,dab1995low,heuer1996collective}, nearly three decades ago. Limited by the computational power and algorithms available back then, they created computer glasses with cooling rates roughly nine orders of magnitude larger than in laboratory settings. They were able to locate only a handful of TLS with the requisite tunnel splittings, necessitating uncontrolled extrapolations.  
The situation then improved incrementally~\cite{DVR99,reinisch2004local,reinisch2005moving,damart2018atomistic}, although the algorithmic ability to simulate glasses which are cooled in an experimentally realistic way has remained out of reach until very recently. This limited greatly our microscopic understanding of the universal anomalous thermal behavior of low-temperature glasses from a computational viewpoint.

In this work we leverage the remarkable ability of the swap Monte Carlo algorithm to produce {\it in silico} amorphous materials with fictive temperatures that range from those found in typical experiments to the significantly slower rates found in recent vapor deposition studies~\cite{ninarello2017models}.  We find a dramatic depletion of active TLS (those with a tunnel splitting $\sim 1 \,$K) with decreasing quench rate, 
as found in recent experiments~\cite{queen2013excess,perezcastaneda2014,liu2014,ediger2014vapor,Queen2015}, with the notable exception of old amber glasses~\cite{perez2014two}.  
We use a state-of-the-art reaction path-finding protocol~\cite{jonsson1998nudged,henkelman2000climbing} to efficiently locate double-well potentials in the multi-dimensional potential energy landscape, yielding a {\em direct sampling} of tunneling states with sufficient statistics to avoid any extrapolation. We determine the degree of localization of individual TLS, providing a detailed, real-space understanding of how atoms participate in tunneling motion and how the thermal exploration of the energy landscape in well-annealed amorphous materials determines the density of tunneling centers. 

 \begin{figure}[t]
   \centering{
  \includegraphics[width=\columnwidth]{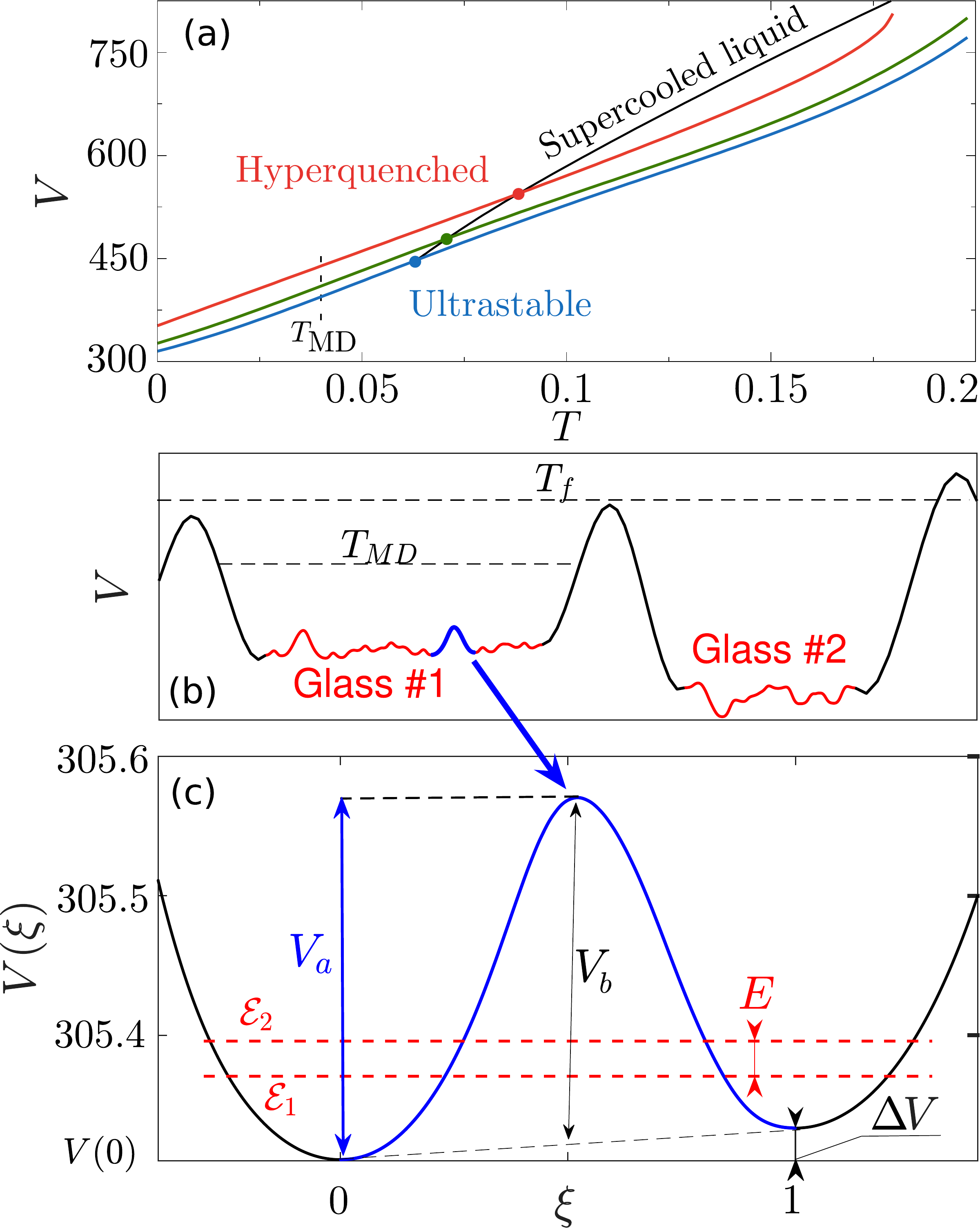}}
  \caption{\label{fig:DWP} (a) Glasses are prepared at equilibrium (black line) at temperatures $T_f = 0.092, 0.07, 0.062$ (bullets), from hyperquenched to ultrastable. We follow their potential energy after rapid quenches (colored lines). (b) Sketch of the potential energy landscape. Double-well potentials are detected with molecular dynamics simulations at $T_{\textrm{MD}} = 0.04$ (blue). (c) A detected double-well potential $V(\xi)$. The classical asymmetry $\D V$, activation energy $V_a$, energy barrier $V_b = V_a - \D V/2$, the energy levels, $\mathcal{E}_1$ and $\mathcal{E}_2$, of the ground-state doublet, and the tunnel splitting, $E = \mathcal{E}_2 - \mathcal{E}_1$, are illustrated.} 
\end{figure} 

{\it Glass preparation -- } 
Past works on the landscape of low-temperature glasses focused on simple models for real materials such as NiP~\cite{heuer1993microscopic},
Argon~\cite{DVR99}, and silica~\cite{reinisch2005moving,damart2018atomistic}. Our goal is to understand how glass preparation affects the density of TLS, within a single model. We study a polydisperse mixture of particles interacting via an inverse power law potential~\cite{ninarello2017models}. Our choice is motivated by the fact that low-temperature anomalies are observed in glasses, regardless the material. Given the diversity of models, we choose one for which swap Monte Carlo enables maximally efficient thermalization on the computer over a range of temperatures at least as wide as in experiments~\cite{ninarello2017models}.

We provide minimal details on the system and measures of equilibration (see~\SI~for details). We simulate a non-additive polydisperse mixture of $N=1500$ particles of mass $m$. Two particles $i$ and $j$ separated by a distance $r_{ij}$ interact via the potential
\begin{equation}
v(r_{ij}) = \epsilon \left(\frac{\sigma_{ij}}{r_{ij}}\right)^{12}+ \epsilon F(r_{ij}/\sigma_{ij})
\label{eq:potential}
\end{equation}
only if $r_{ij} < r_{cut} = 1.25 \sigma_{ij}$, $\sigma_{ij}$ being the non-additive interaction length scale. 
The function $F$ is a fourth-order polynomial which guarantees continuity of the potential up to the second derivative at $r_{cut}$. We characterize the physical classical dynamics of the model using molecular dynamics (MD) with energies and lengths expressed in units of $\epsilon$ and the average diameter $\sigma$, respectively. Times measured during MD simulations are expressed in units of $\sqrt{\epsilon/ m \sigma^2}$.
Number density is set to $\rho=1$.
The relaxation time $\tau_{\alpha}$ of the equilibrium fluid is measured from the self-intermediate scattering function $F_s(k = 7.0, \tau_{\alpha}) = e^{-1}$. 
The onset of glassy dynamics, signaled by deviations from Arrhenius behavior, 
takes place at $T_o = 0.18$, where $\tau_\alpha(T_o) \equiv \tau_o$, and the mode-coupling crossover temperature is $T_{MCT}=0.104$~\cite{ninarello2017models}.

We analyze \textit{in silico} glasses by preparing fully equilibrated configurations at various preparation temperatures $T_f$ using the swap Monte Carlo algorithm, 
which utilizes the exchange of particles' diameters in addition to standard translational moves, leading to a massive thermalization speed-up.
Our implementation is that of Ref.~\cite{berthier2018efficient}.
The configurations are then rapidly cooled to 
lower temperatures using regular molecular dynamics. Therefore, 
$T_f$ represents the ``temperature at which the glass would find itself in equilibrium if suddenly brought to it from its given state,'' which is precisely the definition of Tool's fictive temperature~\cite{tool}. The temperature $T_f$ characterizes the degree of stability of the glasses, see Fig.~\ref{fig:DWP}(a). In experiments, $T_f$ would be determined by the cooling rate~\cite{ediger96,cavagna2009supercooled}, 
or the substrate temperature in a vapor deposition experiment~\cite{swallen2007organic,kearns2008hiking,tylinski2016vapor,perspectiveusg,rafols2017role}. We present results for glasses in wide range of stabilities: poorly annealed (hyperquenched) glasses ($T_f = 0.092$ where $\log(\tau_\alpha/\tau_o)=4.9$, slightly below $T_{MCT}$), liquid-cooled experimental glasses ($T_f = 0.07 \simeq T_g$, where $\log(\tau_\alpha/\tau_o)=10.7$), and ultrastable glasses ($T_f = 0.062$, where $\log(\tau_\alpha/\tau_o)=14.8$). To obtain statistically significant results, we analyze $N_g$ independent samples ($N_g = 200, 50, 15$ for increasing $T_f$).

\begin{figure}[t]
   \includegraphics[width=\columnwidth]{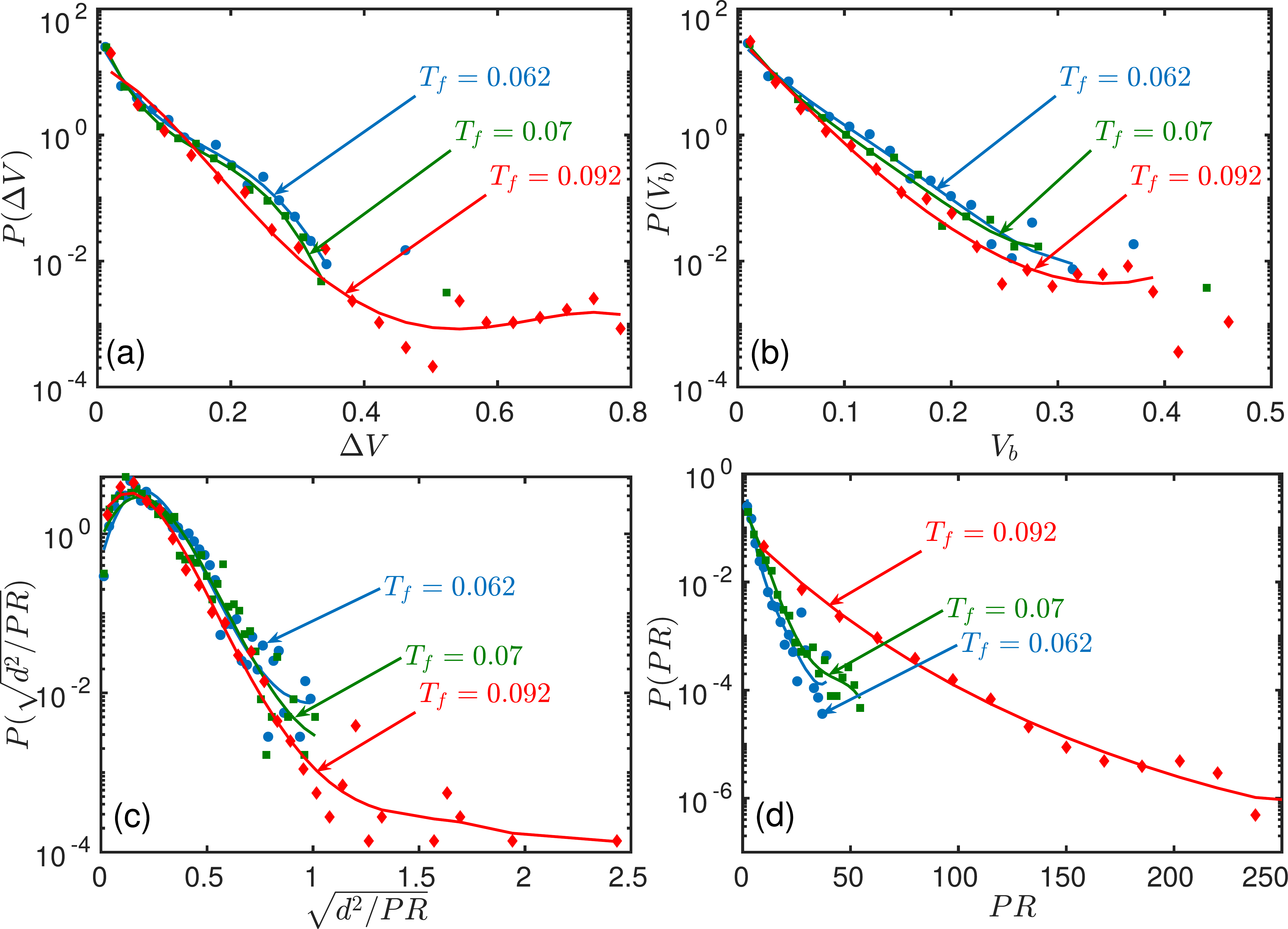}
  \caption{Probability distribution functions of DWP parameters as a function of glass preparation temperature $T_f$: (a)~asymmetry $\Delta V$, 
  (b)~energy barrier $V_b$, (c)~distance $d$ normalized by $\sqrt{PR}$, which characterizes the typical individual displacement of particles that participate actively in a double-well transition, 
  and (d)~participation ratio $PR$. Lines are a guide for the eye.}
  \label{fig:classicalstat} 
\end{figure} 

{\it Landscape exploration -- } 
We identify transitions between nearby minima, or double well potentials (DWPs) in the glasses. Briefly, starting from the configurations equilibrated at $T_f$, we run MD 
simulations at $T_{\textrm{MD}} = 0.04$, which is sufficiently low to confine each glass in a single metabasin, but high enough that the system can rapidly visit distinct minima (inherent structures) within the metabasin~\cite{He08}, see Fig.~\ref{fig:DWP}(b). Details are given in the \SI. 

By sampling the inherent structures along MD trajectories~\cite{DVR99}, we obtain a library of visited minima, as well as the pairs of them that are dynamically connected. We use the isoconfigurational ensemble~\cite{widmer2004reproducible}: for each of the $N_g$ independent configurations, we run up to 200 MD simulations, each initialized with different velocities. The number and duration of isoconfigurational runs is large enough for the probability distribution function (pdf) of the inherent structures potential energy, and the number of transitions, to reach stationarity. While we reach convergence of the pdfs, all possible minima are not sampled, and their numbers increase with additional runs.  
We however sample a significant number of minima: 13252, 26898, 848698 for $T_f = 0.062, 0.07, 0.092$, respectively. As shown below, this is enough to directly determine the density of tunneling TLS.  
 
We select transitions between adjacent minima as described in the \SI. We compute the minimum energy path connecting each pair of minima using a climbing image Nudged Elastic Band (NEB) algorithm~\cite{jonsson1998nudged,henkelman2000climbing}, which ensures the accurate determination of the saddle point, and provides a smooth energy profile. Occasionally, especially for $T_f = 0.092$, the NEB energy profile contains intermediate minima. In such cases, we apply an iterative method to split the energy profile into distinct DWPs, which are then analyzed similarly to the other ones. 

\begin{figure}[t]
   \includegraphics[width=\columnwidth]{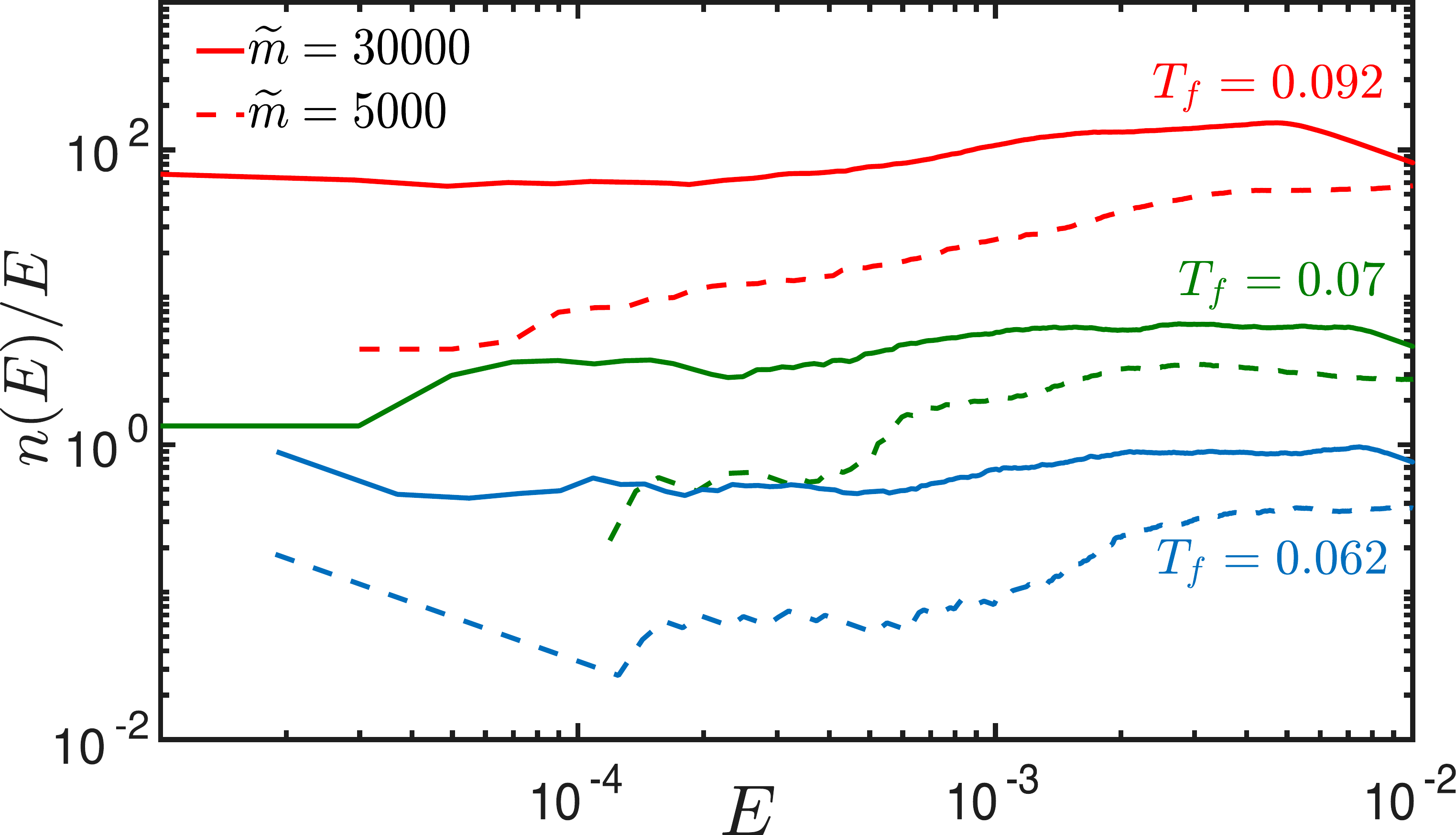}
  \caption{Cumulative distribution of energy splitting $n(E)$ divided by $E$, from hyperquenched to ultrastable glasses (top to bottom). The values $\widetilde m$ are chosen for comparison with real materials. The plateau at small $E$ 
affords a direct determination of the TLS density $n_0$, which is suppressed by two orders of magnitude as glass preparation is varied.}
\label{fig:n0}   
\end{figure} 

We parametrize a DWP by its potential energy $V(\xi)$ along the minimum energy path, with $\xi = 0, 1$ corresponding to the two minima (we arbitrarily choose $\xi =0$ for the deepest minimum), see Fig.~\ref{fig:DWP}(c). A DWP is characterized by its asymmetry $\D V = V(1) - V(0)$, energy barrier $V_b = V_{a} - \D V/2 $, where $V_a$ is the activation energy, and the distance $d$ between minima. The distance is calculated along the reaction coordinate given by the NEB, as the sum of Euclidean distances between images of the system:
$d^2 = \sum_{i,\mu} d_{i,\mu}^2$, where $d_{i,\mu}$ is the displacement of particle $i$ in direction $\mu=x,y,z$. The participation ratio, $PR = d^4/(\sum_{i,\mu} d_{i,\mu}^4)$, characterizes the number of particles involved in the transition. 

We present in Fig.~\ref{fig:classicalstat} the statistics of the DWP parameters. The pdfs for $T_f = 0.062$ and $0.07$ agree quantitatively, within noise, while we observe an evolution for $T_f = 0.092$. In particular, the pdfs of asymmetries and energy barriers are almost exponential in all glasses. The mild dependence of these tails on $T_f$ may stem from the fact that the sampling temperature $T_{\rm MD}$ sets a limit on the DWPs that can be detected, independently of $T_f$ (see \SI~for the effect of $T_{\rm MD}$). While the distribution of energy asymmetry is exponential at high energies $\Delta V$, it becomes flat at low energies
where TLS are found (see Fig. S6). The pdfs of distances (not shown) and participation ratios vary more significantly between $T_f \leq 0.07$ and $T_f = 0.092$. Since $d \propto \sqrt{PR}$, an increase of $PR$ will affect the distribution of $d$. We instead study the pdf of $d/\sqrt{PR}$. This quantity can be interpreted as an average displacement of the particles that participate in the transition. On average, the number of particles involved in DWPs is larger in poorly annealed glasses, while the displacements of individual particles remain comparable. To our knowledge, 
the dependence of the quench rate on DWPs classical parameters has not yet been reported.
Note that TLS typically correspond to DWPs with very small $\D V$ and relatively large $V_b$. The tunnel splitting stems from non-trivial correlations between the classical parameters~\cite{DVR99}, thus the pdfs of classical parameters alone are not informative on quantum tunneling (see \SI).

{\it Density of two-level systems -- } At sufficiently low temperatures, thermal activation over the energy barrier $V_b$ is suppressed, and quantum tunneling becomes important~\cite{gillan1987quantum}. In our analysis, we reduce the problem to a one dimensional effective
Schr\"odinger equation along the reaction coordinate.
Following Vineyard~\cite{vineyard1957frequency}, the effective mass remains $m$, with a reaction coordinate $x\in [0,d]$.
Using the normalized variable $\xi=x/d$, 
and scaling energies by $\epsilon$, we obtain
\begin{equation}
-\frac{\hbar^2}{2 m d^2 \epsilon} \partial^2_\xi \Psi(\xi) + V(\xi)\Psi(\xi) = \mathcal{E}\Psi(\xi),
\label{eq:schro}
\end{equation}
where the ``quantumness'' of the problem is controlled by the 
dimensionless mass 
$\widetilde m=m \sigma^2 \epsilon / \hbar^2$
(see Fig.~\ref{fig:n0}).
In general, the Laplacian should take into account curvature effects, which we neglect here.The potential $V(\xi)$ obtained from the NEB is defined in $\xi \in [0,1]$. To solve Eq.~\eqref{eq:schro}, we extrapolate it outside this interval (see \SI). 

We solve Eq.~\eqref{eq:schro} for all DWPs. We compute the first two energy levels,  $\mathcal{E}_1$ and $\mathcal{E}_2$, of the double well and define the tunnel splitting $ E = \mathcal{E}_2 - \mathcal{E}_1$. We illustrate in Fig.~\ref{fig:DWP}(c) the two energy levels and tunnel splitting of a DWP. The tunnel splitting $ E$ is the relevant parameter for low-temperature properties. 
The transitions that occur by quantum tunneling have a tunnel splitting $ E \sim T$~\cite{phillips87}.
These particular DWPs are called tunneling two-level systems (TLS).

We characterize the distribution of TLS using a cumulative distribution of tunnel splittings $n( E)$,
defined as the number of DWPs with tunnel splitting smaller than $ E$, normalized by the number of particles $N$ in the glass, and the number of independent samples $N_g$. In TLS theory, $n( E)$ can be expanded as $n( E) \simeq n_0  E+ \mathcal{O} ( E ^2)$ for small $E$, the specific heat at low temperature is linear with $T$, and $n_0$ enters the prefactor~\cite{anderson72,phillips87}.

In order to estimate the density $n_0$ of TLS and its dependence on glass stability, we plot $n( E)/ E$ as a function of tunnel splitting $E$ in Fig.~\ref{fig:n0}. All curves indicate a saturation to a plateau value, $n_0$, at low $E$. The existence of a plateau value demonstrates our ability to directly estimate the density of TLS $n_0$ without any uncontrolled extrapolation. Data on tunneling rates~\cite{andreassen2017precision} and distribution of tunneling matrix elements $\Delta_0$ are presented in the SI.

Our key result is that the TLS density $n_0$ (as estimated by $n(E)/E$ in the range $10^{-3}-10^{-4}$) decreases by two orders of magnitude from hyperquenched to ultrastable glasses. To our knowledge, this constitutes the first numerical evidence for a significant suppression of TLS with increasing glass stability. 

\begin{figure}[t]
\includegraphics[width=\columnwidth]{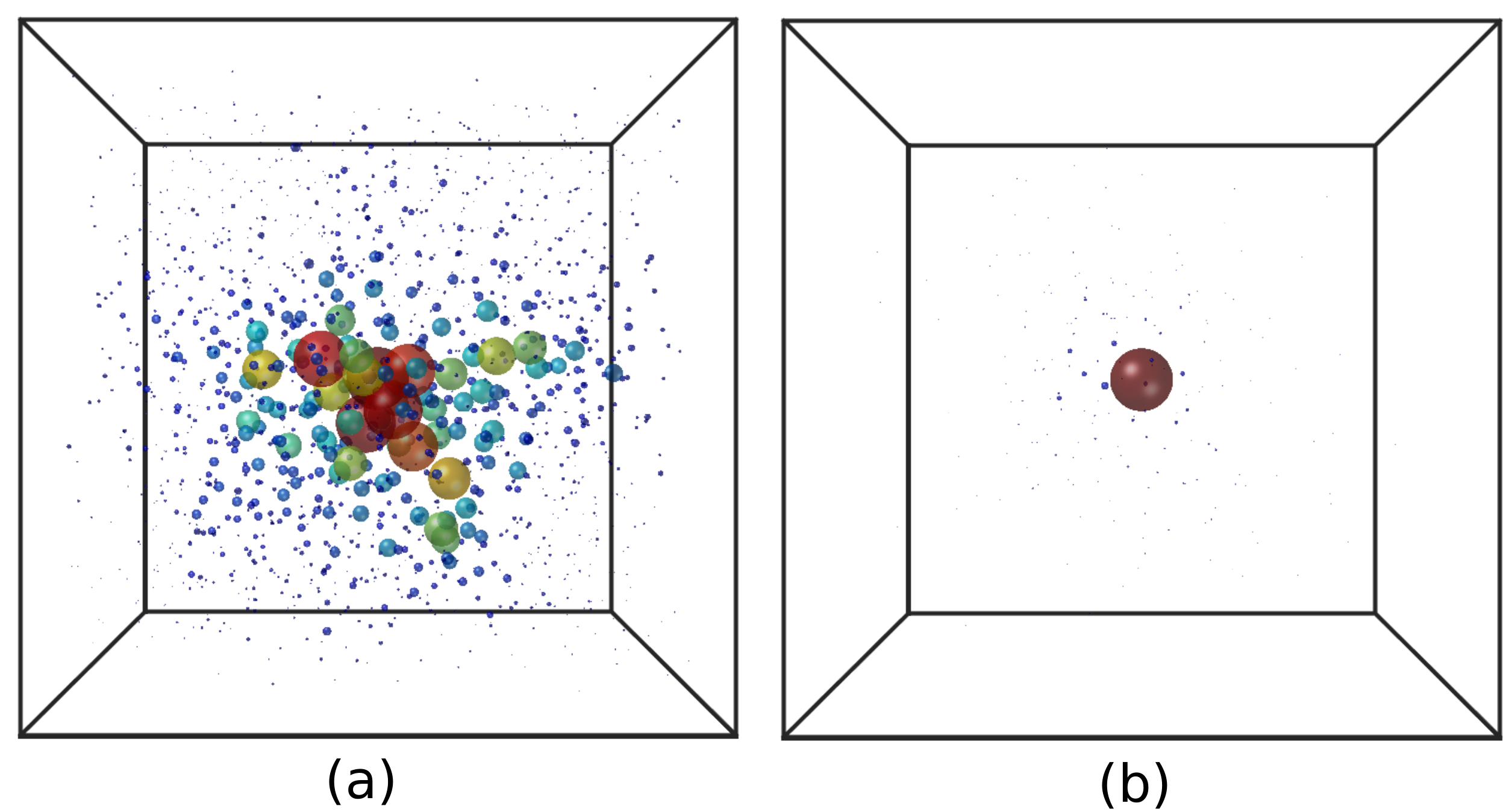}
\caption{Snapshots of TLS with low tunnel splitting $ E$ for $T_f=0.092$ and 
$\widetilde m=30000$. (a) $PR\approx126$ with $ E = 8.9 \times 10^{-5}$. (b) $PR\approx1.6$ and similarly low $ E = 5.4\times 10^{-5}$. The size and color of particles are proportional to their displacement between the initial and final configurations of the TLS, normalized to the highest displacement.}
\label{fig:snapshot}
\end{figure}

 {\it Microscopic nature of TLS -- } How many particles are involved in the tunneling motion of a TLS?~\cite{DVR99,reinisch2004local,reinisch2005moving,damart2018atomistic} 
We analyzed how the participation ratio of transitions correlates with the tunnel splitting $ E$. We find that the participation ratio of low-temperature active TLS with tunnel splittings $ E \sim 10^{-3} - 10^{-4}$ varies from 1 to 200 (see~\SI). The higher participation ratios ($PR \sim 200$) are found in hyperquenched glasses, while in ultrastable glasses the participation ratio rarely exceeds $\sim 30$. We provide systematic numerical evidence that TLS active at low temperature are typically very localized, but occasionally associated with collective excitations. 
We provide two snapshots in Fig.~\ref{fig:snapshot}, corresponding to a collective TLS (left) and a very localized TLS (right) identified in a hyperquenched glass.
 
{\it Discussion -- }Our study of tunneling TLS in a simple computer model demonstrates their importance to understand low-temperature glass anomalies. We show that the density $n_0$ of TLS directly controls the linear temperature dependence of the specific heat at low temperatures. Several recent works advocated the idea that quantized low-frequency harmonic modes alone could explain this behavior~\cite{galperin1989,buchenau1993,parshin1994,parshin2007,wyart2019,franz2019impact,baggioli2019random}. These soft modes are known for our glasses~\cite{wang2019low}, but we find that their contribution to the low-temperature specific heat is subdominant (see \SI), suggesting that the specific heat of structural glasses is dominated by tunneling TLS, as originally proposed in~\cite{anderson72,phillips87}.

To relate our data to experiments we convert simulation units into physical ones. The temperature scale below which quantum effects become important is obtained by comparing the thermal wavelength to the interparticle distance:
$T_Q = \frac{ 2 \pi \hbar^2}{m \sigma^2 k_B}$. We consider DWPs with
$ E < k_B T_Q$ as low-temperature active TLS, and their total number is $n_{active} = N N_g n( E = k_B T_Q)$.
 
A detailed analysis on experimental comparisons is presented in the \SI. We first consider Argon parameters: $\sigma=3.4\times10^{-10}\text{m}$, $\epsilon/4=1.65\times10^{-21}\text{J}$, 
$m= 6\times 10^{-26}$kg~\cite{DVR99}.
This gives $T_g\sim 32 \,$K, $T_Q\sim 0.73\,$K,
and $\widetilde m\sim 4000$. For this choice, we estimate from Fig.~\ref{fig:n0} $n_0^{sim} \sim 4, 0.4, 0.04$ for increasing glass stability. This gives $n_0^{exp}\sim 10^{49}$, $10^{48}$, $10^{47}$ J$^{-1}$m$^{-3}$. 
Active TLS have $ E < k_BT_Q = 0.0015\epsilon$ and 
we find $n_{active} = 1008$, 291, 61 such TLS for $T_g=0.092$, 0.07, 0.062, respectively.
A second choice motivated by NiP metallic glasses~\cite{heuer1993microscopic} would be to use Nickel as a reference, for which
$\sigma=2.21\times10^{-10}\text{m}$, $\epsilon=6.14\times10^{-20}\text{J}$, 
$m=1.02\times 10^{-25}$kg~\cite{ching1977structural}.
In this case, we have $T_g\sim 298\,$K, $T_Q\sim 0.9\,$K,
and $\widetilde m\sim 30000$.
For this value of $\widetilde m$, we find $n_0^{sim} \sim 60, 6, 0.6$ for $T_f=0.092$, 0.07, 0.062, yielding $n_0^{exp}\sim  10^{50}, 10^{49}, 10^{48}$ J$^{-1}$m$^{-3}$. 
Active TLS have $ E < k_B T_Q = 0.0002\epsilon$ and we 
find $n_{active} = 248$, 46, 28 such TLS for $T_g=0.092$, 0.07, 0.062, respectively.

Experimentally, a value of $n_0\sim 10^{46} \text{J}^{-1} \text{m}^{-3}$ is reported~\cite{phillips87,BM88}. Our estimation for $T_f=0.07$ is larger by a factor $\sim 10^2-10^3$ (Argon and Nickel, respectively). 
It is difficult to rationalize this discrepancy but we can offer several hypothesis.
 One possibility is that we include in our estimates all DWPs detected at
temperature $T_{\textrm{MD}}=0.04 \gg T_Q$, while experimentally, the
glass is directly brought to  $T_Q$ and only a small fraction
of TLS that lie at the bottom of the energy metabasin
would be excited. Furthermore, not all TLS would tunnel
on the relevant timescales: it is known that $n_0 \sim \log (\tau)$ where $\tau$ is the observation time at $T_Q$~\cite{phillips87}. This should persist up to the timescale of complete exploration of the energy landscape. Our exploration protocol at $T_{\rm MD} \gg T_Q$ artificially sets $\tau$ larger than this cutoff (see \SI~for the tunneling rates of TLS). Another explanation could be
that our model is too simple to describe real molecular
materials, for example network glasses. Since we analyze
a single model, the fundamental question of universality
in TLS density remains unanswered. Analyzing different glass-forming models will be
crucial to answer this question.

The reduction of $n_0$ by two orders of magnitude when
moving from hyperquenched to ultrastable glasses is
robust and in good agreement with recent experiments~\cite{perezcastaneda2014,Queen2015}. We show that for a given glass-forming model, glass preparation affects dramatically the density of TLS. 
Our results demonstrate that glass ultrastability
(rather than potential anisotropy of the vapor-deposited samples) is responsible for the depletion of TLS.

\vspace{.5cm}

\begin{acknowledgments}
We thank M.~Ediger, D.~Rodney and W.~Ji for discussions. This project received funding from the European Research Council (ERC) under the European Union's Horizon 2020 research and innovation program, Grant No. 723955 -- GlassUniversality (FZ), and Grant No. 740269 --ADNeSP (Prof. M. E. Cates), and from the Simons Foundation (\#454933, LB, \#454955, FZ, \#454951 DR). CS acknowledges the Fondation l'Or\'eal.
\end{acknowledgments}

\bibliography{main.bib}

\end{document}


\title{Supplemental Material: Depletion of two-level systems in ultrastable computer-generated glasses}

\author{Dmytro Khomenko*}

\affiliation{Laboratoire de Physique de l'Ecole Normale Sup\'erieure, ENS, Universit\'e PSL, CNRS, Sorbonne Universit\'e, Universit\'e de Paris, 75005 Paris, France
}

\author{Camille Scalliet*}

\affiliation{
Department of Applied Mathematics and Theoretical Physics, University of Cambridge, Wilberforce Road,
Cambridge CB3 0WA, United Kingdom}

\author{Ludovic Berthier}  

\affiliation{Department of Chemistry, University of Cambridge, Lensfield Road, Cambridge CB2 1EW, United Kingdom}

\affiliation{Laboratoire Charles Coulomb (L2C), Universit\'e de Montpellier, CNRS, 34095 Montpellier, France}

\author{David R. Reichman}

\affiliation{Department of Chemistry, Columbia University, New York, NY 10027, USA}

\author{Francesco Zamponi}  

\affiliation{Laboratoire de Physique de l'Ecole Normale Sup\'erieure, ENS, Universit\'e PSL, CNRS, Sorbonne Universit\'e, Universit\'e de Paris, 75005 Paris, France
}

\date{\today}

\maketitle

\setcounter{equation}{0}
\setcounter{figure}{0}
\setcounter{table}{0}
\setcounter{page}{1}
\renewcommand{\theequation}{S\arabic{equation}}
\renewcommand{\thefigure}{S\arabic{figure}}
\renewcommand{\bibnumfmt}[1]{[S#1]}
\renewcommand{\citenumfont}[1]{S#1}
\renewcommand\thesubsection{\arabic{subsection}}
\renewcommand\thesubsubsection{\arabic{subsection}.\arabic{subsubsection}  }


\section{Model}

We study a three-dimensional, non-additive, continuously polydisperse mixture of particles interacting via the pair potential Eq.~(1) (main text).
The particle diameters $\sigma_i$ are drawn from the normalized distribution $P(0.73 <\sigma< 1.62) \sim 1/\sigma^3$, with average diameter $\langle \sigma \rangle= \int P(\sigma) \mathrm{d} \sigma = 1$ (simply denoted as $\sigma$ in the main text). We employ a non-additive cross-diameter $\sigma_{ij} = 0.5 (\sigma_i+\sigma_j)\left( 1-0.2|\sigma_i-\sigma_j|\right)$. This model is efficiently simulated 
with the swap Monte Carlo algorithm. The choice of particle dispersity and non-additivity make the homogeneous fluid robust against fractionation and crystallization at all temperatures numerically explorable~\cite{ninarello2017models}. As in the main text, classical quantities are given in units
of $\epsilon, \sigma, m$.

We characterize the dynamic slowdown of the supercooled liquid using molecular dynamics (MD) simulations. The temperature evolution of the relaxation time $\tau_{\alpha}$ provides three temperatures relevant to this work. The onset of glassy dynamics, signaled by deviations from Arrhenius behavior, takes place at $T_o = 0.18$, where $\tau_{\alpha} (T_o) = \tau_o$. The mode-coupling crossover temperature $T_{MCT} = 0.104$ is measured by fitting the relaxation time data to $\tau_{\alpha} \sim (T - T_{MCT})^{-\gamma}$ in a limited window of relatively high temperatures. At $T_{MCT}$, the dynamics is four orders of magnitude slower than at the onset temperature, $\tau_{\alpha} (T_{MCT}) \simeq 10^4 \tau_o$. We take $T_{MCT}$ as a proxy for the computer glass transition, below which standard MD fail to equilibrate the homogeneous fluid. The experimental glass transition $T_g = 0.067$ is located by fitting the relaxation time data to a parabolic law, using $\tau_{\alpha} (T_g) = 10^{12} \tau_o$~\cite{ninarello2017models}. 

This glass-forming model is efficiently simulated at equilibrium with the swap Monte Carlo algorithm. We employ the swap Monte Carlo algorithm implemented in the LAMMPS package, with the optimal parameters, as described in Ref.~\cite{berthier2018efficient}. Supercooled liquid configurations can be generated down to $T=0.062$, i.e. below the experimental glass transition $T_g$. In this work, we focus on configurations prepared in equilibrium conditions at temperatures $T_f = 0.092, 0.07, 0.062$. 
For the two lowest $T_f$ values, standard MD dynamics initialized from an equilibrium configuration is completely arrested: no diffusion is observed and the system is trapped within a glass metabasin. For the higher $T_f=0.092$, some diffusion is observed in equilibrium; but if the system is rapidly cooled at lower temperatures, once again no diffusion is observed and a glass state is obtained.
Borrowing from experimental conventions~\cite{tool}, 
we call this the ``fictive temperature'' of the glasses, as discussed in the main text.

\section{Landscape exploration}

The amorphous configurations generated at $T_f$ are first thermalized to $T_{\textrm{MD}} = 0.04$ using a Berendsen thermostat. At this temperature, no diffusion is observed for all glasses.
The system is then simulated in the NVE ensemble, using an integration time step of $dt=0.01$. Configurations along the MD trajectory are used as the starting point for energy minimization via a conjugate gradient algorithm, which brings them to their inherent structure (IS), i.e. the nearest local minimum in the potential energy landscape (PEL). Note that contrary to the MD simulations, the minimization procedure does not conserve the total energy. We minimize the MD configurations obtained every 20, 10, 5 time steps for $T_f = 0.062, 0.07, 0.092$. The high frequency of minimization is chosen to identify nearby local minima, separated by an energy barrier. 

From the MD simulations, we obtain a time series of inherent structures. We are interested in transitions between two different inherent structures, identified by comparing two consecutive minima. We recorded 70970, 130859, and 1593359 transitions between inherent structures, for $T_f = 0.062, 0.07, 0.092$ respectively. We wish to characterize the potential energy barriers corresponding to the transitions. This analysis is computationally costly. Given the large number of transitions detected, we chose to analyze transitions detected at least 4 times, regardless of the direction of the transition ($A \rightarrow B$ or $B \rightarrow A$). The number of transitions analyzed is equal to 14195, 23535, and 117361, for $T_f = 0.062, 0.07, 0.092$ respectively.

\begin{figure}[t]
 \includegraphics[width=0.7\columnwidth]{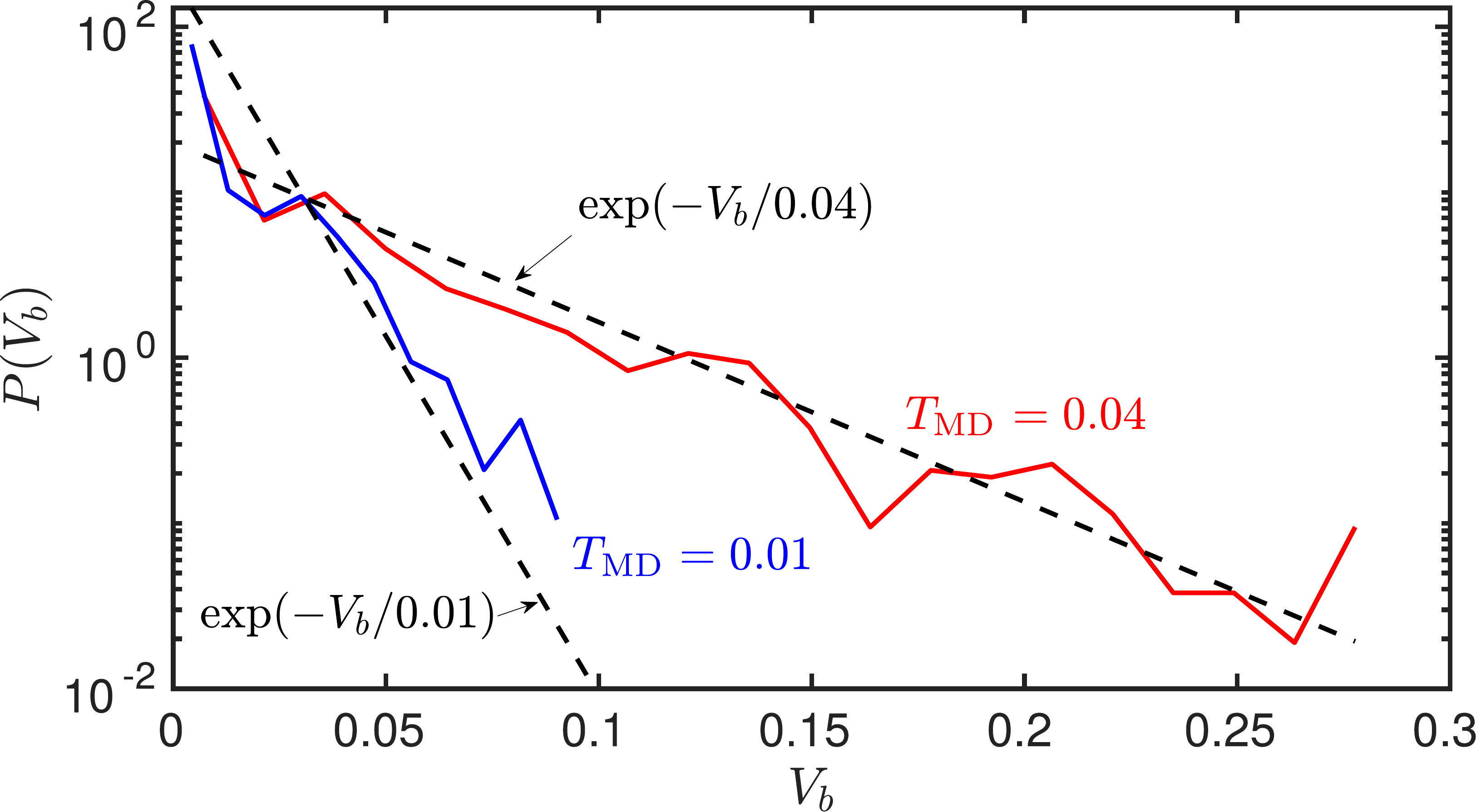}
 \caption{Probability distribution of the potential energy barriers $V_b$ sampled at temperatures $T_{\textrm{MD}} = 0.01$ and $0.04$. } 
 \label{fig:SIpv}
\end{figure}

In order to investigate the influence of the temperature $T_{\textrm{MD}}$ on the characteristics of the double well potentials identified, we present in Fig.~\ref{fig:SIpv} the probability distribution function of potential energy barriers identified with $T_{\textrm{MD}} = 0.01$, and $0.04$ 
(used in the main text). The sampling temperature $T_{\textrm{MD}}$ influences the tail of the distribution only, which decays as $\exp(-V_b/T_{\textrm{MD}})$. We conclude that the sampling of relevant TLS is not
affected by a variation of $T_{\rm MD}$ within a reasonable interval.

\section{Tunnel splitting}

For each analyzed transition, the Nudged Elastic Band (NEB) algorithm outputs a one-dimensional potential defined for the reduced coordinate $\xi$, between the two minima only. We run the algorithm using 40 images of the system. We need to extrapolate the potential in order to obtain a full double well potential. We tested various extrapolation schemes, such as parabolic fitting of the minima, and mirroring around each minima, defining $V(-\xi) = V(\xi)$ for $\xi<\xi_a$ and $V(2-\xi) = V(\xi)$ for $\xi>\xi_a$, where $\xi_a$ is the coordinate of the potential maximum: $V(\xi_a)=V(0)+V_a$. In the main text, we used a linear extrapolation of the reaction path obtained with the NEB. Let us denote $\mathbf{r}_1$ and $\mathbf{r}_2$ the coordinates of the particles in the first two images of the system along the reaction path ($\mathbf{r}_1$ is an energy minimum). We measure the potential energy of the configuration, starting from $\mathbf{r}_1$, and moving in the direction $\mathbf{r}_1 - \mathbf{r}_2$. We perform a similar extrapolation at the other minimum. We show in Fig.~1 (main text) a double-well potential obtained from the NEB algorithm (blue part), and by linear extrapolation of the reaction path (black part). We have compared all methods and found that while each scheme gives a slightly different potential, the statistics of tunnel splittings remains unaffected by our choice. In the main text, we use the most physical scheme, namely the linear extrapolation of the reaction path. 

Once the classical potential $V(\xi)$ is obtained by extrapolation as discussed above, we solve
numerically the Schr\"odinger Eq.~(2) (main text) using a standard Python package. Note that,
in general, the Laplacian term should take into account curvature effects along the reaction coordinate,
$\nabla^2_\xi=\partial^2_\xi+(\text{det} g)^{-1/2}\partial_{\xi} [(\text{det} g)^{1/2}\partial_i]+(\text{det} g)^{-1/2}\partial_{i} [(\text{det} g)^{1/2}\partial_\xi]$, where $g$ is a metric tensor and $\xi_i$ are coordinates orthogonal to $\xi$. For simplicity, we neglect these effects and use the standard second derivative along the reaction coordinate.

\begin{figure}[t]
    \includegraphics[width=\columnwidth]{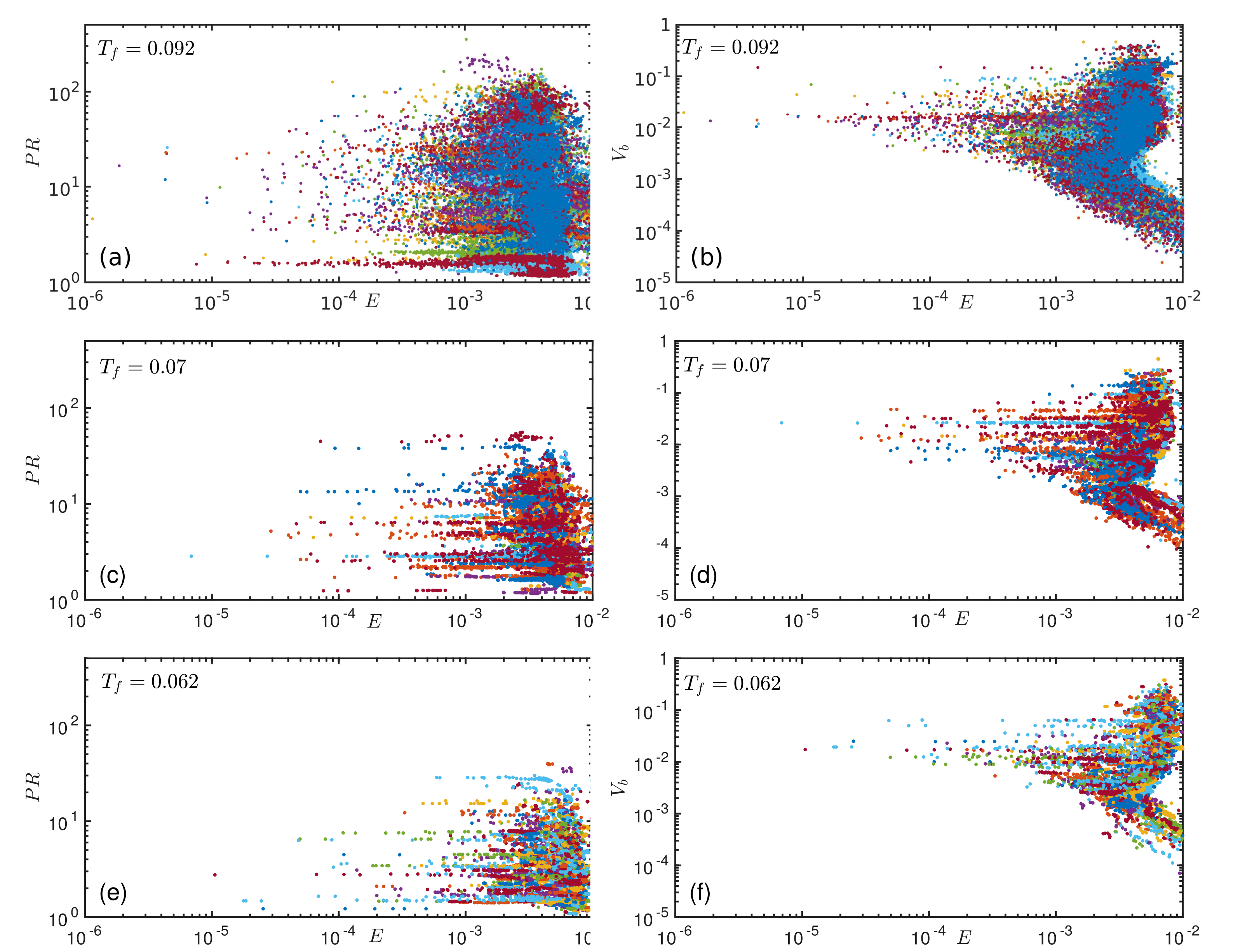}
 \caption{Scatter plot of the participation ratio $PR$ (a,c,e) and the potential energy barrier $V_b$ (b,d,f) 
versus the tunnel splitting $ E$,
 of double-well potentials in glasses prepared at $T_f = 0.092$ (a-b), $T_f = 0.07$ (c-d), and $T_f = 0.062$ (e-f). The data for DWPs found in the same glass sample (there are $N_g$ of them) are presented with the same color. The tunnel splittings are computed using an adimensional mass $\widetilde m = 30000$. 
 }
 \label{fig:scatpr} 
\end{figure}

 We present in Fig.~\ref{fig:scatpr} a scatter plot of the tunnel splitting $ E$ as a function of participation ratio $PR$ and energy barrier $V_b$ for DWPs identified in glasses from hyper-quenched (top) to ultrastable (bottom). In Fig.~\ref{fig:scatpr}~(a,c,e) we observe that in the relevant range of $ E \lesssim 10^{-3}$ the value of $PR$ can be as large as $\sim 200$ for $T_f=0.092$ and $\sim 30$ for $T_f=0.062$. 
 In Fig.~\ref{fig:scatpr}~(b,d,f) we observe that in the same relevant range, the barrier is always $V_b\gtrsim 10^{-2} \gg E$, indicating that the relevant DWP are indeed TLS. We also checked (not shown) that in the same range,
 one always has $\mathcal{E}_3 - \mathcal{E}_1 \gg  E$, where $\mathcal{E}_3$ is the third energy level.
 
 Finally, we checked that for most TLS in the relevant range, the wavefunctions
 of the first two levels are sufficiently delocalized over the two wells, indicating that tunnelling is active. However, the barrier $V_b$ has a relatively wide distribution (see Fig.~\ref{fig:scatpr}), and there exist TLS with large barriers and hence very small tunnelling matrix elements. For those TLS, the wavefunctions are almost localized, indicating that tunnelling is highly suppressed. These TLS would be frozen in experimental conditions. The wide distribution of $V_b$ is known to be responsible for a logarithmic growth of $n_0$ with observation time $\tau$, $n_0 \propto \log \tau$~\cite{phillips87}, which might also provide an explanation for why our simulations overestimate $n_0$ with respect to experiments (see main text).
  
\begin{figure}[t]
  \includegraphics[width=0.7\columnwidth]{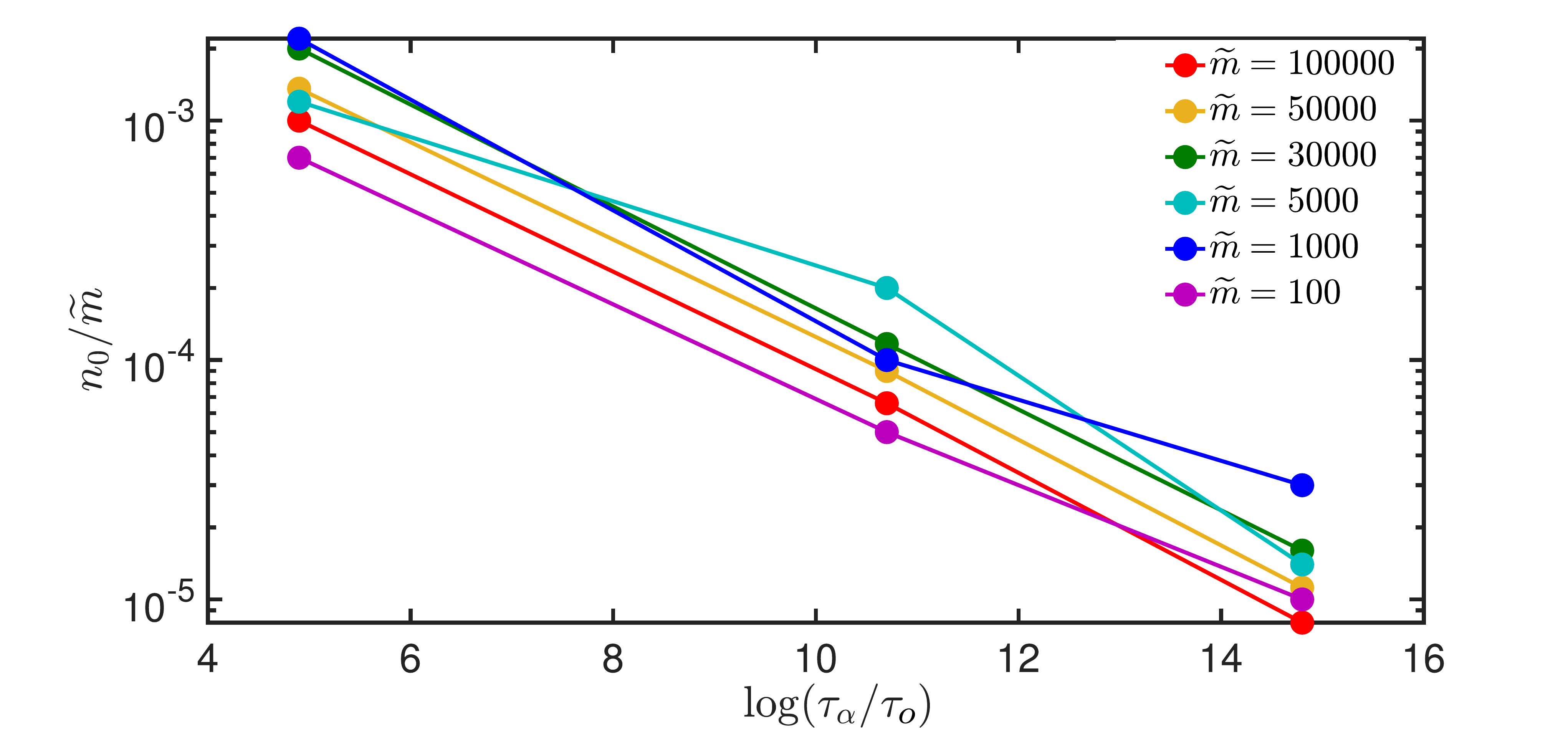}
 \caption{\label{fig:n0_m0} 
The TLS density $n_0$, as extrapolated from $n(E)/E$ in the limit $E\to 0$, divided
by $\widetilde m$, as a function of glass stability, encoded by the value of $\log(\tau_\alpha/\tau_o)$ for the three values of $T_f$ and different 
$\widetilde m$. A reduction of $n_0$ by
two orders of magnitude is robustly observed, independently of $\widetilde m$. Furthermore,
we find $n_0 \propto \widetilde m$ independently of $T_f$.
 }
\end{figure} 

\section{Dimensional scaling analysis for unit conversion}

The number of TLS per particle in a given glass sample with a tunnel splitting
less than $ E$, for $ E\to 0$, is $n_0  E$. Hence, $n_0$ has the dimensions of an inverse energy, expressed in units of $\epsilon^{-1}$. 
However, because in our simulation units the number density is $\rho=1$, $n_0$ is also the number of TLS per unit volume, in units
of $\epsilon^{-1} \sigma^{-3}$, i.e.
\begin{equation}
    n_0^{exp} = n_0^{sim} \times \epsilon^{-1} \sigma^{-3}\ .
\end{equation}
For Argon, 
$\epsilon^{-1}\sigma^{-3} \sim 3.85\times 10^{48} \text{J}^{-1} \text{m}^{-3}$,
while for Nickel $\epsilon^{-1}\sigma^{-3} \sim 1.51\times 10^{48} \text{J}^{-1} \text{m}^{-3}$,
which allows one to convert our numerical results for $n_0^{sim}$ into experimental values for these
two materials.

Because the glass transition temperature in the simulation is $T_g = 0.067$,
the corresponding glass transition temperature in physical units is
\begin{equation}
    T_g = 0.067 \times \frac{\epsilon}{k_B}
    \ .
\end{equation}
Finally, 
the temperature $T_Q$ at which quantum effects become relevant
is that at which the thermal wavelength equals the interparticle distance. Since our simulation density
is $\rho=1/\sigma^3$, we get
\begin{equation}
  T_Q = \frac{ 2 \pi \hbar^2}{m \sigma^2 k_B}
  = \frac{2\pi}{\widetilde m} \frac{\epsilon}{k_B}
    \ ,
\end{equation}
and we recall that 
$\widetilde m=m \sigma^2 \epsilon / \hbar^2$
is the dimensionless mass that appears
in the Schr\"odinger equation.
Finally, note that the relevant (active) TLS are those with $ E < k_B T_Q$,
and their total number in our simulation is given by 
\begin{equation}\begin{split}
n_{active} &= N N_g n\left( E = k_B T_Q \right)\\ 
&\sim N N_g n^{sim}_0 \frac{k_B T_Q}{\epsilon}=2\pi N N_g \frac{n^{sim}_0}{\widetilde m} \ .
\end{split}\end{equation}
We find that $n_0/\widetilde m$ is roughly constant for a given glass stability $T_f$, as shown in Fig.~\ref{fig:n0_m0}. This results implies that the choice of $\widetilde m$ can be quite arbitrary,
within a reasonable range. In particular, $\widetilde m$ cannot be too small otherwise the condition
$T_Q \ll T_g$ would be violated.

\section{Vibrational and TLS contributions to the specific heat}

Here, we compare the contributions to specific heat coming from TLS and from harmonic vibrations around an inherent structure.
To obtain the TLS contribution, for a given $T_f$ and $\widetilde m$, we collect all the TLS found in all the $N_g$ glasses, with splitting
$ E_i$, and compute~\cite{phillips87,anderson72}
\begin{equation}
C_{\rm TLS} = 
\frac{1}{N N_g} \sum_{i} \frac{x_i^2}{\cosh(x_i)^2}\ ,
\qquad x_i = \frac{ E_i}{2 k_B T} \ .
\end{equation}

To obtain the vibrational contribution, we considered a single
glass prepared at $T_f=0.062$, and all the $N_{\rm IS}$ inherent structures found within that glass. For each inherent structure, we diagonalize the Hessian matrix of the classical potential, to obtain a set of eigenvalues (spring constants) $\kappa_\alpha$. Each of these provides a quantum harmonic oscillator contribution to the vibrational specific heat, which is
\begin{equation}
    C_{\rm vib} = \frac{1}{N N_{\rm IS}} \sum_\alpha \frac{x_\alpha^2}{\sinh(x_\alpha)^2} \ ,
    \qquad x_\alpha = \frac{\hbar \omega_\alpha}{2 k_B T} \ ,
    \qquad \omega_\alpha =\sqrt{\frac{\kappa_\alpha}{m}} \ .
\end{equation}
Note that the finite size of the system imposes a cutoff
on the low-frequency Debye behavior. In fact, the lowest
frequency found in our system is
$\hbar \omega_\alpha / k_B = 0.0038$ (in simulation units) 
for $\widetilde m=5000$, 
and $\hbar \omega_\alpha / k_B = 0.0016$ for $\widetilde m=30000$.
To avoid this problem we used data of larger system with $N=192000$ particles~\cite{wang2019low} and extrapolate the quadratic region of the density of states to the $\omega\to 0$ limit.

\begin{figure}[t]
 \includegraphics[width=\columnwidth]{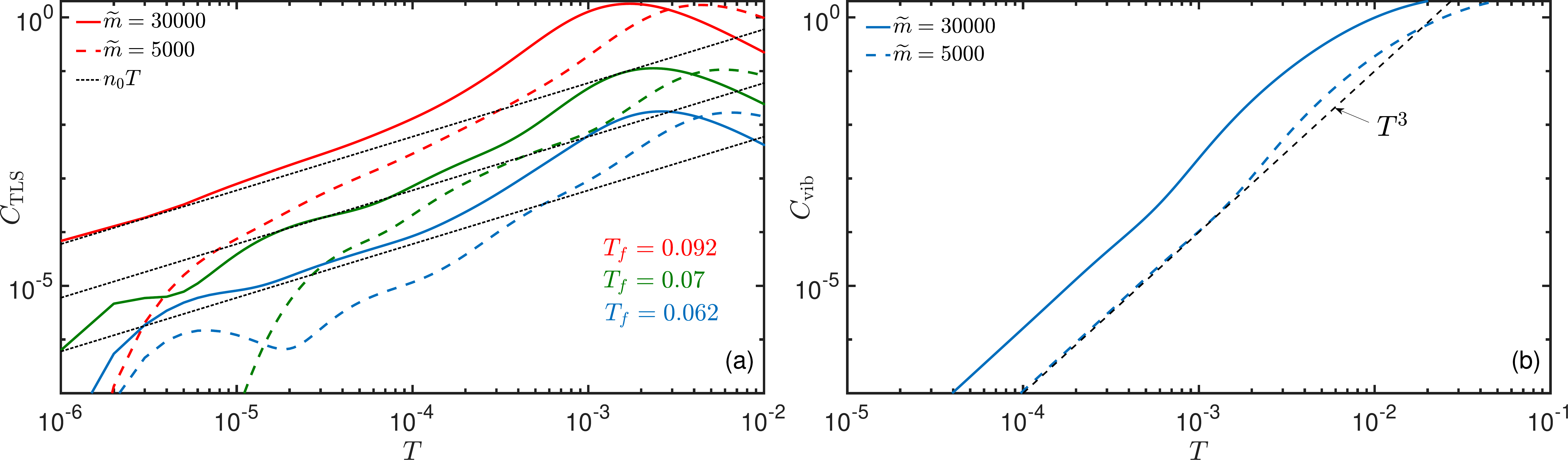}
 \caption{\label{fig:C} 
(a) TLS contribution to the specific heat per particle, as a function of $T$. Dotted-dashed lines are the asymptotic low-temperature scaling, $C_{\rm TLS} \sim T$, for $\widetilde m=30000$ (for $\widetilde m=5000$ the statistics is not sufficient for a good extrapolation). (b)~Vibrational contribution to the
 specific heat per particle, as a function of $T$, for $T_f=0.062$, averaged over inherent structures in a typical glass metabasin. }
\end{figure}

Yet, from Fig.~\ref{fig:C} we conclude that the vibrational contribution decays much faster than the TLS one upon lowering temperature, and that already at $T\sim 10^{-4}$ we have
$C_{\rm vib}\ll C_{\rm TLS}$ for all considered values of
$\widetilde m$ and $T_f$. 

\section{Tunneling rates}

\begin{figure}[t]
 \includegraphics[width=\columnwidth]{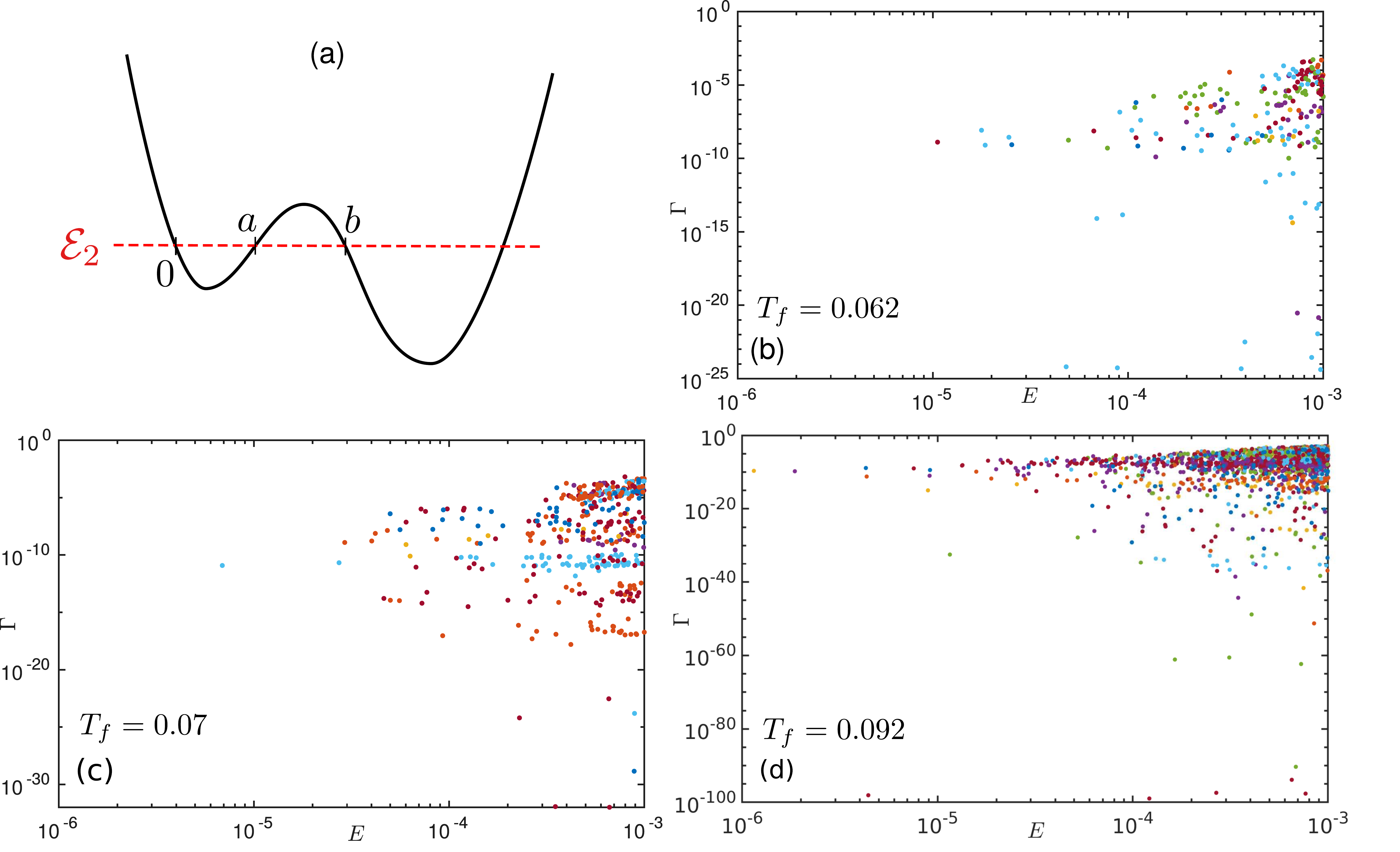}
 \caption{(a) Definition of the parameters $a$ and $b$ used in the WKB calculation of tunneling rates, Eq.~\eqref{eq:rate}. (b-d) Transition rates $\Gamma$ as a function of energy splitting $E$ for different glass stabilities, $T_f = 0.062$ (b), 0.07 (c), 0.092 (d). The data for DWPs found in the same glass sample (there are $N_g$ of them) are presented with the same color. The rates are computed with NiP parameters ($\widetilde m=30000$).}
 \label{fig:rate} 
\end{figure} 

In this section we compute the tunneling rates of the TLS identified in our simulations. This allows us to determine if the TLS with energy splitting $E \lesssim k_B T_Q$, which we considered as ``active'' at $T\lesssim T_Q$, can actually tunnel over experimental timescales.

The characteristic time to observe a transition from the higher energy level to the lower one can be estimated from the decay rate $\Gamma$, computed within the Wentzel-Kramers-Brillouin (WKB) approximation~\cite{andreassen2017precision}
\begin{equation}
    \Gamma=\frac{1}{m}\left[\int_0^a\frac{dx}{p(x)}\right]^{-1}\exp\left[-\frac{2}{\hbar}\int_a^b|p(x)|dx\right], \quad \quad  p(x)=\sqrt{2m(\mathcal{E}_2-V(x))},
    \label{eq:rate}
\end{equation}
where $[0, a]$ and  $[a, b]$ are the width of the higher energy well
 and of the energy barrier, respectively, both measured at the energy level $\mathcal{E}_2$. This construction is sketched in Fig.~\ref{fig:rate}(a).
We present in Fig.~\ref{fig:rate}(b-c) the scatter plots of the tunneling rate $\Gamma$ of a DWP as a function of its energy splitting $E$. Each panel corresponds to a different glass stability, from ultrastable (b) to hyperquenched (d).

To compare these results to experimental timescales, we consider the parameters for NiP ($\widetilde m=30000$), for which $\epsilon = 6.14 \times 10^{-20}$ J. Together with $\hbar = 1.05 \times 10^{-34}$ m$^2$kg/s, we get a natural frequency unit $\epsilon/\hbar \sim 10^{14}$ Hz for $\Gamma$. Most TLS with $E < 10^{-3}$, which could tunnel around $T_Q$, have rates $\Gamma \geq 10^{-14}$, which corresponds to $\Gamma \geq 1$ Hz in physical units.
This means that most TLS identified as active based on their energy splitting would actually tunnel over experimental timescales.

\section{Distribution of tunneling matrix element $\Delta_0$}

\begin{figure}[h]
 \includegraphics[width=\columnwidth]{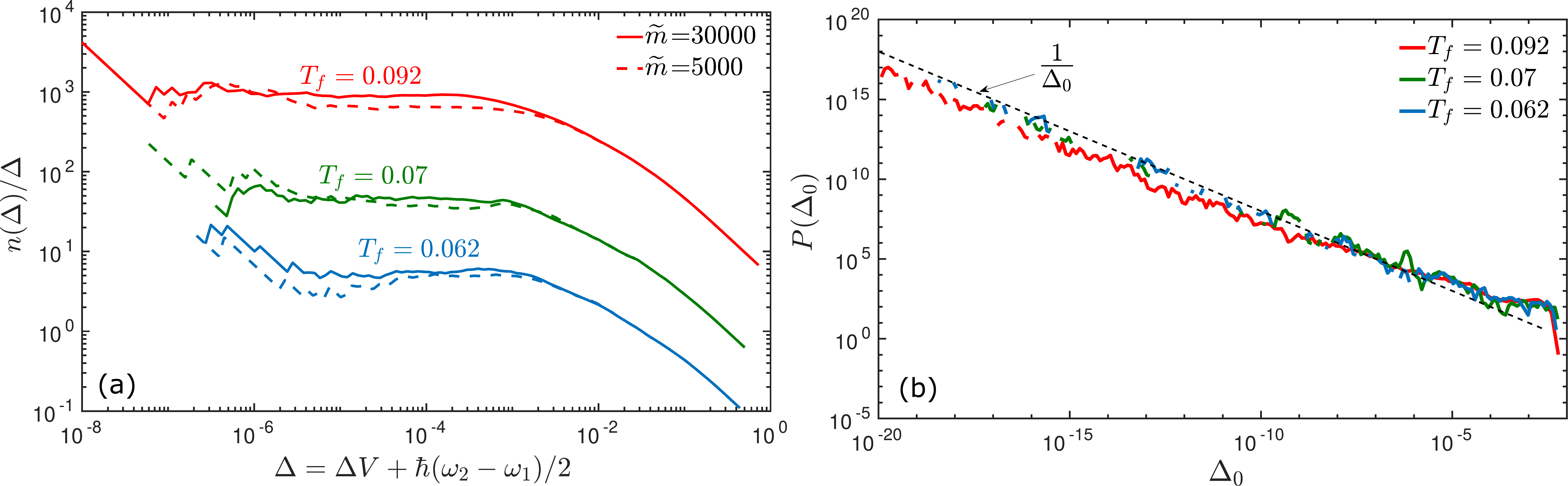}
 \caption{(a) Cumulative distribution of the diagonal splitting $\Delta$, divided by $\Delta$ for different glass stabilites. The data is shown in a way similar to Fig. 3 of the main article. (b) Probability distribution of the off-diagonal element $\Delta_0$ for different glass stabilities. Our results are consistent with the TLS model prediction $P(\Delta_0) \sim 1/\Delta_0$. Data obtained for $\widetilde m=30000$ (NiP parameters).}
 \label{fig:delta0} 
\end{figure}

In the standard TLS model, the Hamiltonian of a single tunneling state takes the form
\begin{equation}
H = \frac{1}{2}\begin{pmatrix}
\Delta & \Delta_0 \\
\Delta_0 & -\Delta 
\end{pmatrix}
\end{equation}
in the localized representation. 
The diagonal splitting of the TLS can be estimated by 
\begin{equation}
\Delta = \Delta V + \hbar \frac{\omega_2 - \omega_1}2 \ ,
\end{equation}
where $\omega_{1,2}$ are the classical frequencies in the two energy minima~\cite{anderson72}.
The tunneling matrix element is $\Delta_0=2\langle\psi_L|H|\psi_R\rangle$, where $|\psi_{R,L}\rangle$ are the wave functions of the left and right wells. While this definition has been widely used in the theoretical literature, it is not obvious how to define the localized basis $|\psi_{R,L}\rangle$, and therefore not practical to obtain a numerical evaluation of $\Delta_0$. 
We used instead the WKB approximation to estimate the tunneling matrix element~\cite{phillips87} 
\begin{equation}\label{d0}
    \Delta_0\approx \overline{\cal E}\exp\left[-\frac{1}{\hbar}\int_a^b|p(x)|dx\right]  \ ,
    \qquad \overline{\cal E} = \frac{{\cal E}_1 + {\cal E}_2}2 \ ,
\end{equation}
where here $a$, $b$, and $p(x)$ are derived from $\overline{\cal E}$ instead of ${\cal E}_2$ as in the calculation of $\Gamma$.
Note that the most important contribution, in both cases, comes from the exponential term.
We show in Fig.~\ref{fig:delta0}(a) the resulting cumulative density of diagonal elements $\Delta$ (in the same format as Fig.~3 in the main text), and
probability distribution function $P(\Delta_0)$ of tunneling elements, computed for each $T_f$. Our numerical results are consistent with the TLS model prediction $P(\Delta_0) \sim 1/\Delta_0$. The density of diagonal elements is constant at small $\Delta$, also in agreement with the TLS model. However, the condition $\Delta_0 \ll \Delta$ is not always satisfied
in our data, and as a result the true splitting $E$ is generally larger than $\Delta$, which results in a decrease of the plateau values of $n(E)/E$ compared to
$n(\Delta)/\Delta$.

\bibliography{tls.bib}